\documentstyle[prd,aps,epsfig,twocolumn,floats]{revtex}
\begin{document}
\draft
% \wideabs{
\title{The transformations of non-abelian gauge fields
under translations}
\author{Bernd A. Berg (berg@hep.fsu.edu)}
\address{Department of Physics, The Florida State University,
  Tallahassee, FL 32306}
\date{June 6, 2000 -- hep-th/0006045, revised June 12, 2000}
% \date{printed \today}
\maketitle
\begin{abstract}
I consider infinitesimal translations
$x'^{\alpha}=x^{\alpha}+\delta x^{\alpha}$
and demand that Noether's approach gives a symmetric 
energy-momentum tensor as it is required for gravitational sources.
This argument determines the transformations of non-abelian gauge
fields under infinitesimal translations to differ from the usually
assumed invariance by the gauge transformation, $A'^a_{\,\ \gamma} (x') 
 - A^a_{\,\ \gamma}(x) =
\partial_{\gamma} [ \delta x_{\beta}\, A^{a\,\beta}(x)] + C^a_{\,\ bc}\, 
\delta x_{\beta}\, A^{c\,\beta}(x)\, A^{b}_{\,\ \gamma}(x)$
where the $C^a_{\,\ bc}$ are the structure constants of the gauge group. 

\end{abstract}
%}

\narrowtext

\vskip 10pt
% \section{Introduction} \label{sec:intro}

In a previous paper~\cite{bb_em} I have determined the transformations
of the electromagnetic potentials under translations from the requirement
that the energy-momentum tensor as it comes out of
Noether's theorem~\cite{Noether,BS} ought to be symmetric. Such a result
is desireable, because the energy-momentum tensor enters as source
of the gravitational field and the symmetry transformations of
general covariance yield a symmetric tensor, see for
instance~\cite{LaLi02,WeiG}. A more detailed motivation is given in
my first paper. Here I extend the argument to non-abelian gauge
theories.

Following the notation of Weinberg's book~\cite{WeiII}, the
Lagrange density is
\begin{equation} \label{Lgauge}
 {\cal L} = - {1\over 4}\, F^a_{\,\ \alpha\beta}\, F^{a\,\alpha\beta}
\end{equation}
with
\begin{equation} \label{Fgauge}
 F^a_{\,\ \alpha\beta} = \partial_{\alpha} A^a_{\,\ \beta} 
  - \partial_{\beta} A^a_{\,\ \alpha} +
  C^a_{\ bc}\, A^b_{\,\ \alpha}\, A^c_{\,\ \beta} 
\end{equation}
where the $C^a_{\ bc}$ are the structure constants of the gauge
group. The Lagrangian~(\ref{Lgauge}) is invariant under the gauge
transformations of the fields:
\begin{equation} \label{gauge_t}
 A^a_{\,\ \alpha} \mapsto A^a_{\,\ \alpha} + \partial_{\alpha}
 \epsilon^a (x) + C^a_{\ bc}\, \epsilon^c (x)\, A^b_{\,\ \alpha}\ .
\end{equation}
As in~\cite{bb_em} the observation is that the conventionally assumed
invariance of the gauge fields under translations
\begin{equation} \label{translations}
x'^{\, \alpha} = x^{\alpha} + \delta x^{\alpha}
\end{equation}
can be enlarged by gauge transformations and the general form reads
\begin{equation} \label{gauge_trans_ansatz}
 A'^a_{\,\ \gamma} (x') = A^a_{\,\ \gamma}(x) +
 \partial_{\gamma} \epsilon^a (x) + C^a_{\ bc}\, \epsilon^c (x)\,
 A^b_{\,\ \gamma}\ .
\end{equation}
Repeating the arguments of Noether's theorem in the version of~\cite{BS}
and requesting a symmetric energy-momentum tensor determines the gauge
transformation uniquely and leads to the transformation law stated in
the abstract
\begin{eqnarray} \nonumber
A'^a_{\,\ \gamma} (x') & = & A^a_{\,\ \gamma}(x) + \partial_{\gamma}\,
[ \delta x_{\beta}\, A^{a\,\beta}(x) ] \\  \label{gauge_trans}
& + & C^a_{\,\ bc}\, \delta x_{\beta}\,
A^{c\,\beta}(x)\, A^{b}_{\,\ \gamma}(x)\ .
\end{eqnarray}
The remainder of this letter is devoted to the derivation of this
equation and my treatment follows closely~\cite{bb_em} where also
a few additional steps can be found.

First, let us consider general fields $\psi_k$ and recall the
derivation of the relativistic Euler Lagrange equations 
from the action principle. The action is a four dimensional 
integral over a scalar Lagrangian density
\begin{equation} \label{action}
{\cal A} = \int d^4x\, {\cal L} 
(\psi_k,\, \partial_{\alpha}\psi_k)\ .
\end{equation}
Variations of the fields are defined as functions
\begin{equation} \label{fvariations}
\delta\psi_k (x) = \psi'_k (x) - \psi_k (x)\, 
\end{equation}
which are non-zero for some localized space-time region. The action
is required to vanish under such variations
$$ 0 = \delta {\cal A} = $$
\begin{equation} \label{Lvar1}
 \sum_k \int d^4x\, \left[
(\delta\psi_k)\, {\partial {\cal L}\over \partial \psi_k} + 
(\delta\, \partial_{\alpha}\psi_k)\, {\partial {\cal L}\over 
\partial (\partial_{\alpha}\psi_k)} \right]\ . 
\end{equation}
Integration by parts allows to factor $\delta \psi_k$ out and,
because all the $\delta \psi_k$ are independent, we arrive at
the Euler-Lagrange equations
\begin{equation} \label{EulerLagrange}
{\partial {\cal L}\over \partial \psi_k} - \partial_{\alpha}\,
{\partial {\cal L}\over \partial (\partial_{\alpha}\psi_k)} = 0\ .
\end{equation}
Together with the anti-symmetry of $F^a_{\,\ \alpha\beta}$ in the
Lorentz indices, the Euler-Lagrange equations imply the relation
\begin{equation} \label{ELrelation}
 \partial_{\gamma}\, {\partial {\cal L}\over \partial
 (\partial_{\alpha} A^a_{\,\ \gamma})} = 
  C^b_{\ ac}\, A^c_{\,\ \beta}\, {\partial {\cal L}\over \partial
 (\partial_{\beta} A^b_{\,\ \alpha})}\ .
\end{equation}

Noether's theorem applies to transformations of the 
coordinates for which the transformations of the field functions are
also known and we introduce, in addition to (\ref{fvariations}), a
second type of variations which combines space-time and their
corresponding field variations
\begin{equation} \label{cvariations}
\overline{\delta} \psi_k (x) = \psi'_k (x') - \psi_k (x)\, . 
\end{equation}
Using
$$ \psi'_k (x') = \psi'_k (x) + 
\delta x^{\alpha}\ \partial_{\alpha} \psi_k (x) $$
we find a relation between the variations (\ref{cvariations}) and 
(\ref{fvariations})
\begin{equation} \label{rvar}
\overline{\delta} \psi_k (x) = \delta\psi_k (x) +
\delta x^{\alpha}\ \partial_{\alpha} \psi_k (x)\, .
\end{equation}
For a scalar field $\psi$ symmetry under translations means
\begin{equation} \label{ctrans}
\overline{\delta} \psi (x) = \psi' (x') - \psi (x) = 0\, . 
\end{equation}
But for the gauge fields we allow~(\ref{gauge_trans_ansatz})
\begin{eqnarray} \nonumber
\overline{\delta} A^a_{\,\ \gamma} (x) & = & A'^a_{\,\ \gamma} (x') -
 A^a_{\,\ \gamma} (x) \\ \label{vtrans}
 & = & \partial_{\gamma} \epsilon^a (x) +
       C^a_{\ bc}\, \epsilon^c (x)\, A^b_{\,\ \gamma} (x)
\end{eqnarray}
and equation (\ref{rvar}) becomes
\begin{equation} \label{rvtrans}
\delta A^a_{\,\ \gamma} = \partial_{\gamma} \epsilon^a (x) +
  C^a_{\ bc}\, \epsilon^c (x)\, A^b_{\,\ \gamma}
- \delta x^{\alpha}\ \partial_{\alpha} A^a_{\,\ \gamma} (x)\ .
\end{equation}
As the Lagrange density is a scalar, we get for its combined
variation~(\ref{cvariations})
\begin{equation} \label{cvarL}
0 = \overline{\delta} {\cal L} = {\cal L}' (x') - {\cal L} (x) = 
\delta {\cal L}+ \delta x^{\alpha}\ \partial_{\alpha} {\cal L}
\end{equation}
where besides~(\ref{ctrans}) we used the relation~(\ref{rvar}).
Our aim is to factor an over-all variation $\delta x^{\alpha}$ 
out. For $\delta {\cal L}$ we proceed as in equation~(\ref{Lvar1}),
where the $\psi_k$ fields are now replaced by the gauge
fields $A^a_{\,\ \gamma}$
$$ \delta {\cal L} =  (\delta A^a_{\,\ \gamma})\, {\partial 
{\cal L} \over \partial A^a_{\,\ \gamma}} + (\delta \partial_{\alpha} 
A^a_{\,\ \gamma})\, {\partial {\cal L} \over \partial 
(\partial_{\alpha} A^a_{\,\ \gamma})}\,.$$
Using the Euler-Lagrange equation (\ref{EulerLagrange}) to
eliminate $\partial {\cal L} / \partial A^a_{\,\ \gamma}$, we get
$$ \delta {\cal L} = \partial_{\alpha}
 \left[ (\delta A^a_{\,\ \gamma})\, {\partial {\cal L}
 \over \partial (\partial_{\alpha} A^a_{\,\ \gamma})} \right]\, .$$
Let us collect all terms which contribute  to
$\overline{\delta}{\cal L}$ in equation~(\ref{cvarL}).
We find  (note that $\partial_{\beta}\delta x^{\alpha}=0$ holds for
all combinations of indices $\alpha$, $\beta$)
$$ 0=\overline{\delta} {\cal L} = \partial_{\alpha}\, 
\left[ (\delta A^a_{\,\ \gamma})\, {\partial {\cal L}\over \partial 
(\partial_{\alpha} A^a_{\,\ \gamma)}} + \delta x^{\alpha}\ {\cal L}
\right] = $$
$$ \partial_{\alpha}\, \left[ ( \partial_{\gamma} \epsilon^a (x) +
  C^a_{\ bc}\, \epsilon^c (x)\, A^b_{\,\ \gamma} )\, {\partial
  {\cal L}\over \partial (\partial_{\alpha} A^a_{\,\ \gamma})}
 \right]  $$
$$ + \delta x_{\beta}\, \partial_{\alpha}\, \left[
 - ( \partial^{\beta} A^a_{\,\ \gamma} )\, {\partial {\cal L}
   \over \partial (\partial_{\alpha} A^a_{\,\ \gamma})}
 + g^{\alpha\beta}\, {\cal L} \right]  $$
where equation~(\ref{rvtrans}) was used. To be able to factor
$\delta x_{\beta}$ also out of the first bracket on the
right-hand side, one has to request
\begin{equation} \label{translation_gauge}
 \epsilon^a (x) = \delta x_{\beta}\, B^{a\,\beta} (x)
\end{equation}
where $B^{a\, \beta}(x)$ is a not yet determined gauge field. With
this we get
\begin{eqnarray} \label{delta_x}
 0 & = & \delta x_{\beta}\, \partial_{\alpha}\, \left[
 (\partial^{\beta}\,A^a_{\,\ \gamma})\, {\partial {\cal L} \over
 \partial (\partial_{\alpha} A^a_{\,\ \gamma})} \right. \\ \nonumber
& - & \left. ( \partial_{\gamma}\, B^{a\,\beta}
+ C^a_{\ bc}\, B^{c\,\beta}\, A^b_{\,\ \gamma} )\,
 {\partial {\cal L} \over \partial (\partial_{\alpha} A^a_{\,\ \gamma})}
- g^{\alpha\beta}\, {\cal L} \right]\,.
\end{eqnarray}
Equation~(\ref{ELrelation}) implies that the contribution from the
gauge transformations is a total divergence,
\begin{eqnarray} \nonumber
 \partial_{\gamma}\, \left( {\partial {\cal L}\over \partial 
 (\partial_{\alpha} A^a_{\,\ \gamma})}\, B^{a\,\beta} \right) & = & 
\\ {\partial {\cal L}\over \partial 
   (\partial_{\alpha} A^a_{\,\ \gamma})}\,
    \left[ \, \partial_{\gamma}\, B^{a\,\beta} \right.
& + & \left. C^a_{\ bc}\, B^{c\,\beta}\, A^b_{\,\ \gamma}\, \right]\ .
\end{eqnarray}
As the variations $\delta x_{\beta}$ in~(\ref{delta_x}) are
independent, the energy-momentum tensor 
\begin{eqnarray} \nonumber
 \theta^{\alpha\beta} & = & {\partial {\cal L}\over \partial 
 (\partial_{\alpha} A^a_{\,\ \gamma})}\,
 \left[ \partial^{\beta} A^a_{\,\ \gamma} -
 ( \partial_{\gamma}\, B^{a\,\beta}
 + C^a_{\ bc}\, B^{c\,\beta}\, A^b_{\,\ \gamma} ) \right] \\
 & - & g^{\alpha\beta}\, {\cal L}  \label{mtensor}
\end{eqnarray}
gives the conserved currents
\begin{equation} \label{e-mcurrent}
\partial_{\alpha}\, \theta^{\alpha\beta} = 0\, .
\end{equation}
We demand that $\theta^{\alpha\beta}$ is symmetric. The Lagrangian
term $g^{\alpha\beta} {\cal L}$ is manifestly symmetric and we have
to deal with the other contributions. We note that
$$ {\partial {\cal L} \over \partial (\partial_{\alpha}
   A^a_{\,\ \gamma})} = - F^{a\, \alpha\gamma} $$
with $F^{a\, \alpha\gamma}$ given by equation~(\ref{Fgauge}).
Therefore, the choice 
\begin{equation}
B^{a\, \beta}(x) = A^{a\, \beta}(x)
\end{equation}
leads to 
 \begin{equation} \label{stensor}
\theta^{\alpha\beta} = F^{a\, \alpha\gamma}\,
F^{a\,~~\beta}_{\ \ \gamma} - g^{\alpha\beta}\,{\cal L}
\end{equation}
where we used the anti-symmetry of the structure constant under
interchange of $b$ and $c$. The tensor~(\ref{stensor}) is symmetric
because of
$$ F^{a\, \alpha\gamma}\, F^{a\,~~\beta}_{\ \ \gamma} = 
   F^{a\,\beta\gamma}\, F^{a\,~~\alpha}_{\ ~\gamma}\ .  $$

In conclusion, I have derived the transformation
behavior~(\ref{gauge_trans}) by demanding that the energy-momentum
tensor from Noether's theorem comes out symmetric. To the many arguments
why gauge invariance is needed, this adds another one: It is needed
to make the energy-momentum distribution under local translational
variations symmetric.
\hfil\break

\centerline{\bf Note added} \medskip

After posting this manuscript Prof. Jackiw kindly informed
me that my result is a special case of his work~\cite{Ja78},
see~\cite{JaMa80} for details. Prof. Hehl communicated that the
use of 1-Forms leads directly to a symmetric energy-momentum
tensor, see for instance~\cite{He95}.

\acknowledgments

This work was in part supported by the US Department of Energy under
contract DE-FG02-97ER41022.


\begin{thebibliography}{99}

\bibitem{bb_em} B.A. Berg, hep-th/0006029 (unpublished). 

\bibitem{Noether} E. Noether, Nachrichten Akademie der Wissenschaften,
G\"ottingen p.235 (1918).

\bibitem{BS} N.N. Bogoliubov and D.V. Shirkov, 
{\it Introduction to the Theory of Quantized Fields}, 
John Wiley \& Sons, 1959: Chapter 1.

\bibitem{LaLi02} L.D. Landau and E.M. Lifshitz, {\it The Classical
Theory of Fields}, Course of Theoretical Physics~2, Butterworth
Heinemann, Oxford, 1995: Section 94.

\bibitem{WeiG} S. Weinberg, {\it Gravitation and Cosmology}, John
Wiley \& Sons, 1972: Chapter 12.

\bibitem{WeiII}  S. Weinberg, {\it The Quantum Theory of Fields II},
Cambridge University Press, 1998: Chapter 15.

\bibitem{Ja78} R. Jackiw, Phys. Rev. Lett. 41, 1635 (1978).

\bibitem{JaMa80} R. Jackiw and N.S. Manton, Ann. Phys. 127, 257 (1980).

\bibitem{He95} F.W. Hehl, J.D. McCrea, E.W. Mielke, and Y. Ne'eman,
Phys. Rep. 258, 1 (1995).

\end{thebibliography}
\end{document}